\documentclass[12pt,a4paper]{article}
\usepackage{amsmath,amsfonts}
\usepackage[font={footnotesize,it}]{caption}
\usepackage[top=1.2in, bottom=1.2in, left=1.2in, right=1.2in]{geometry}

\DeclareCaptionStyle{italic}[justification=centering]{labelfont={bf},textfont={it},labelsep=colon}
\usepackage{graphicx,psfrag,epsf}
\usepackage{enumerate}
\usepackage{natbib}
\usepackage{url, setspace}
\usepackage{booktabs, subfig, bm, paralist,mathpazo,tikz,todonotes,longtable,microtype,dsfont,rotating}
\usepackage[pdftex,colorlinks=true]{hyperref}
\definecolor{darkblue}{rgb}{0,0,.6}
\hypersetup{citecolor=darkblue,linkcolor=darkblue,urlcolor=darkblue}

\newcommand{\blind}{0}

\newcommand{\Rlogo}{\protect\includegraphics[height=1.8ex,keepaspectratio]{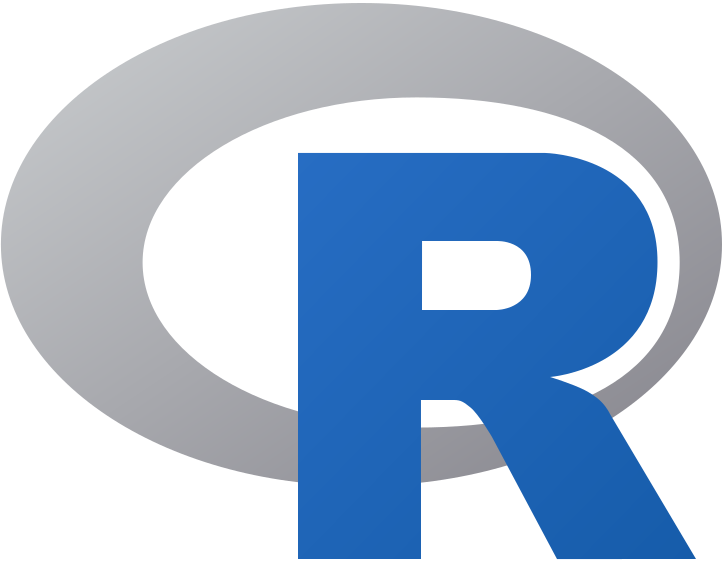}}

\newcommand{\X}{\mathcal{X}}
\newcommand{\Y}{\mathcal{Y}}
\graphicspath{{plots/}}
\DeclareMathOperator*{\argmin}{\arg\!\min}

\newsavebox\CBox

\usepackage{orcidlink}

\date{}

\begin{document}

\def\spacingset#1{\renewcommand{\baselinestretch}%
{#1}\small\normalsize} \spacingset{1}


\if0\blind
{
  \title{\bf Stopping time detection of wood panel compression: A functional time series approach}
  \author{Han Lin Shang \orcidlink{0000-0003-1769-6430} \hspace{.2cm}\\
 Department of Actuarial Studies and Business Analytics\\ Macquarie University\\
\\    Jiguo Cao \orcidlink{0000-0001-7417-6330} \\
    Department of Statistics and Actuarial Science\\ Simon Fraser University \\
\\     Peijun Sang \orcidlink{0000-0002-5312-0204} \\
     Department of Statistics and Actuarial Science\\ University of Waterloo   }
  \maketitle
} \fi

\if1\blind
{
  \title{\bf Stopping time detection of wood panel compression: A functional time series approach}
} \fi

\bigskip
\begin{abstract}
We consider determining the optimal stopping time for the glue curing of wood panels in an automatic process environment. Using the near-infrared spectroscopy technology to monitor the manufacturing process ensures substantial savings in energy and time. We collect a time series of curves from a near-infrared spectrum probe consisting of 72 spectra and aim to detect an optimal stopping time. We propose an estimation procedure to determine the optimal stopping time of wood panel compression and the estimation uncertainty associated with the estimated stopping time. Our method first divides the entire data set into a training sample and a testing sample, then iteratively computes integrated squared forecast errors based on the testing sample. We then apply a structural break detection method with one breakpoint to determine an estimated optimal stopping time from a univariate time series of the integrated squared forecast errors. We also investigate the finite-sample performance of the proposed method via a series of simulation studies.
\end{abstract}

\noindent%
\vfill

\newpage
\spacingset{1.5} 

\section{Introduction}\label{sec:intro}

Functional time series consist of random functions observed over time. Each function, denoted by $\{\X_t(u), t\in \mathbb{Z}\}$, is a realization of a stochastic process $\X(u)$ where $u\in \mathcal{I}\subset \mathbb{R}$ represents a continuum bounded within a finite interval $\mathcal{I}$ that is a subset of the real line $\mathbb{R}$. There has been a surge of interest in studying functional time series that take values in an infinite-dimensional space in recent years. Examples of functional time series include intraday stock price curves with each functional observation defined as a pricing function of time points within a day \citep[e.g.,][]{HKR14}, and age-specific fertility rate curves with each functional observation defined as a function of different ages for a particular calendar year \citep[e.g.,][]{LRS19}. In our wood panel compression data, the continuum ranges near-infrared (NIR) spectrum wavelengths.

The analysis of NIR curves has been attracting extensive attention in chemometrics and less so in statistics. However, in statistics, \cite{Goutis98} considered a scalar-on-function regression with a roughness penalty imposed on the second-order derivative of a functional variable and applied the scalar-on-function regression to predict protein content from a set of wheat spectra. In a scalar-on-function regression, \cite{BFV01} considered a wavelet transformation of the discretized curves and applied a Bayesian variable selection method to the multivariate regression of predictands on wavelet coefficients. Further, they applied the scalar-on-function regression to predict fat, sugar, and water content from the NIR spectrum of biscuit doughs. \cite{FV02} and \cite{Ferraty14} considered a nonparametric framework to capture the relationship between a scalar-valued response and a function-valued predictor. They applied the nonparametric regression to predict the fat content of a meat sample based on its NIR absorbance spectrum. Further, \cite{FV03} extended the nonparametric framework from regression to curve discrimination and classified those spectra that have the fat content larger than 20\% from a set of NIR spectrum of meat samples. In a scalar-on-function regression, \cite{HS21} found that the second derivative of the NIR curves is the optimal covariate for analyzing the fat content.

In the existing literature, the NIR spectra curves were independent and identically distributed (i.i.d.) and were often considered an explanatory variable in a regression that involves at least a functional variable. There is yet work on analyzing a functional time series of NIR curves to the best of our knowledge, with an application to wood panel compression in the lumber industry. We aim to fill the gap by contributing a novel methodology for determining the optimal stopping time of wood panel compression. Our developed method is of practical importance to the lumber industry worldwide, potentially saving manufacturing and labor costs. In particular, we address an important problem from FPInnovations (\url{https://web.fpinnovations.ca}), a not-for-profit R\&D private organization that focuses on creating solutions to boost the growth of the Canadian forest section. As advocated by FPInnovations, using the NIR spectroscopy technology to monitor the manufacturing process ensures substantial savings in energy and time. It increases productivity, thus enhancing the overall competitiveness of the Canadian wood processing industry. Finally, the proposed methodology can be applied to other densely observed functional time series where spectral signals can be extracted to reflect the dynamic process of interest. 

Subtle changes in experimental conditions, such as temperature and pressure in a laboratory, and moisture content in wood, are expected to impact the spectroscopy curves. Ideally, we would like to observe multiple functional time series of NIR spectra and find an optimal stopping time for each of the functional time series. Due to the data availability, we focus on a single functional time series. With a time series of 72 spectra acquired from a near-infrared spectrum probe, we apply a functional time series forecasting method to estimate the optimal stopping time of wood panel compression, along with the estimation uncertainty of the stopping time. 

Our method first uses functional principal component analysis for the functional time series to reduce dimensionality. Secondly, we model each set of principal component scores using an autoregressive integrated moving average to obtain the forecast scores. By multiplying the forecast scores with functional principal components, we obtain forecast curves. We iteratively compute integrated squared forecast errors (ISFEs) of holdout functional time series. Based on a univariate time series of the forecasting errors, we then apply a simple structure break procedure to determine a breakpoint. 

Our work is aligned with many papers in functional change-point analysis literature. There have been many attempts to detect one or more change points and the corresponding different regimes. When there is an abrupt change, it is commonly referred to as the ``change point" for the point in time where the change takes place. The change occurs when a stochastic process exhibits a shift in mean, variance, or distribution. \cite{BGH+09} and \cite{AGH+09} developed statistical methods to test the null hypothesis of no structural break against the alternative of a single break in the mean function, assuming that the errors are i.i.d. curves. By reducing the infinite dimension of the functional time series to a finite dimension via the classic functional principal component analysis, change-point detection methods developed for multivariate time series can be applied to identify breaks \citep[see, e.g.,][]{AGH+09, BGH+09, AK12, Torgovitski15}. \cite{ZSH+11} studied a structural break detection method for serially correlated functional time series, and the method is based on a self-normalization approach of \cite{SZ10}. In a time domain, \cite{ARS18} considered a fully functional approach for detecting structural breaks in functional time series data without dimension reduction. The fully functional approach avoids possible information loss caused by the dimension reduction, thus may have more reliable numerical performance. In a frequency domain, \cite{AV19} considered smooth deviations from stationarity of functional time series. When there is a smoothly changing situation, it is commonly referred to as a ``regime switch" for a different regime after the change point. The detection of regime switch can be thought of as change-point detection with one change point. This is the focus of our paper. Unlike those existing papers, our proposed method considers a functional change-point problem from a forecasting perspective. A standard change point method can be applied to a univariate time series of forecasting errors. 

The remainder of this paper is structured as follows. In Section~\ref{sec:2}, we describe the motivating data set consisting of the NIR spectrum of wood panels. In Section~\ref{sec:3}, we introduce a functional principal component analysis for reducing the dimensionality of a functional time series. In Section~\ref{sec:4}, we introduce a functional time series forecasting method and implement a forecasting scheme to determine forecasting errors, such as ISFEs, for some holdout samples. In Section~\ref{sec:5}, we present a sieve bootstrap method to obtain bootstrapped functional time series forecasts, from which we obtain a distribution of optimal stopping time to quantify uncertainty. In Section~\ref{sec:6}, we present a simulation study. In Section~\ref{sec:7}, we present the results for the motivating data set. From the ISFEs, we apply a structural break detection method to identify the optimal stopping time. Conclusions are drawn in Section~\ref{sec:8}, along with some ideas on how the methodology can be further extended.

\section{Wood panel NIR spectra}\label{sec:2}

This data set consists of 72 sets of spectra acquired from a NIR probe inserted into the wood panel matrix before panel pressing. Spectra were automatically collected during the panel processing. All spectra were acquired using an Analytical Spectral Devices field portable spectrometer. The acquisition range was from 350 to 2300nm, with a 1nm spectral resolution (i.e., 1951 points per spectrum). The actual time of acquisition for each spectrum is embedded. Therefore, the time interval between spectra can be computed as the difference between consecutive spectra times.

\begin{figure}[!htbp]
\centering
\includegraphics[width=10cm]{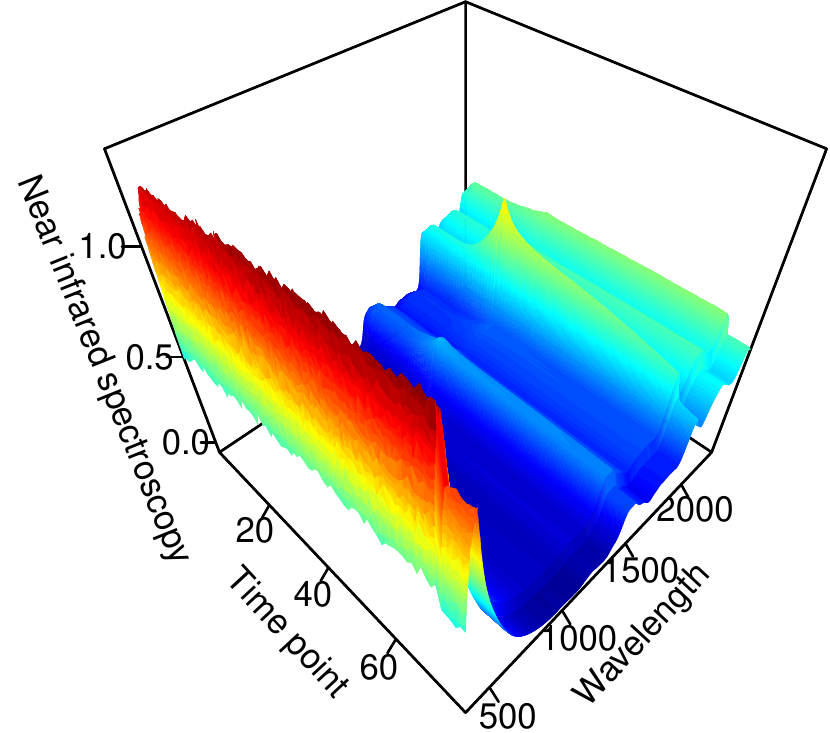}
\caption{A perspective plot of 72 NIR spectra curves with the wavelength from 350 to 2300nm.}\label{fig:1}
\end{figure}

In Figure~\ref{fig:1}, we present a perspective plot of NIR spectra curves. NIR spectra can be viewed as a set of dense functional data. While the perspective plot does not reveal the optimal stopping time, it shows a temporal change in the NIR spectra curves at different wavelengths. The optimal stopping time is one among 72 time points where the wood panel compression completes. We aim to determine the earlier termination time.

\section{A functional time series forecasting method}\label{sec:3}

We do not restrict our considerations to a particular functional time series forecasting method. Many of the functional predictors applied in the statistical literature fit our stopping time detection algorithm in Section~\ref{sec:4}. We elaborate on some examples:
\begin{inparaenum}
\item[1)] functional autoregressive of order one \citep{Bosq00}, where a functional predictor can be lagged observations of a functional response;
\item[2)] nonparametric functional regression \citep{APS06};
\item[3)] functional principal component regression where scores are modeled and forecast via a multivariate time series forecasting method \citep{ANH15}; and 
\item[4)] functional principal component regression where scores are modeled and forecast via a univariate time series forecasting method \citep{HS09}.
\end{inparaenum}
We consider a functional principal component regression where scores are modeled and forecast via a univariate time series forecasting method. The univariate time series forecasting method can model a possible non-stationary series of scores.

\subsection{Functional principal component analysis}

From the sample $\bm{\X}(u) = \{\X_1(u), \dots, \X_n(u)\}$, we compute the sample mean function and sample covariance functions, which are defined by
\begin{align*}
\overline{\X}(u) &= \frac{1}{n}\sum^n_{t=1}\X_t(u), \\
\widehat{\mathcal{C}}(u, v) &= \frac{1}{n}\sum^n_{t=1}\{[\X_t(u) - \overline{\X}(u)][\X_t(v) - \overline{\X}(v)]\}.
\end{align*}
Via Mercer's lemma, the empirical covariance function can be decomposed as
\begin{equation*}
\widehat{\mathcal{C}}(u, v) = \sum^{\infty}_{k=1}\widehat{\lambda}_k\widehat{\phi}_k(u)\widehat{\phi}_k(v),
\end{equation*}
where $\widehat{\lambda}_1>\widehat{\lambda}_2>\cdots\geq 0$ are the sample eigenvalues, and $[\widehat{\phi}_1(t), \widehat{\phi}_2(t), \dots]$ are the corresponding orthogonal sample eigenfunctions. The realizations of the stochastic process $\X$ can be written as
\begin{equation}
\X_t(u) =  \overline{\X}(u) + \sum^{\infty}_{k=1}\widehat{\beta}_{t,k}\widehat{\phi}_k(u), \qquad t=1,2,\dots,n, \label{eq:FPCA_infinite}
\end{equation}
where $\widehat{\beta}_{t,k}$ is the $k$\textsuperscript{th} estimated principal component score for the $t$\textsuperscript{th} time period.

In practice, we truncate~\eqref{eq:FPCA_infinite} into the first $K<n$ terms and a model residual term $\zeta_t(u)$. This leads to
\begin{equation}
\X_t(u) = \overline{\X}(u) + \sum^{K}_{k=1}\widehat{\beta}_{t,k}\widehat{\phi}_k(u) + \zeta_{t}(u), \label{eq:FPCA_finite}
\end{equation}
where expansion~\eqref{eq:FPCA_finite} facilitates dimension reduction as the first $K$ terms often provide a good approximation to the infinite sums, and thus the information contained in $\bm{\X}(u)$ can be properly summarized by the long vector $(\widehat{\bm{\beta}}_1^{\top},\dots,\widehat{\bm{\beta}}_K^{\top})^{\top}$, where $\widehat{\bm{\beta}}_k = (\widehat{\beta}_{1,k}, \ldots, \widehat{\beta}_{n,k})^{\top}$ and $^{\top}$ denotes matrix transpose.

When a functional data set is i.i.d., the functional principal component analysis is an adequate dimension-reduction technique. However, in the presence of temporal dependence, long-run covariance plays an essential role in modeling functional time series \citep{RS17}. The long-run covariance can be defined as
\begin{equation*}
C(u, v) = \sum^{\infty}_{\ell=-\infty}\gamma_{\ell}(u,v),
\end{equation*}
where $\gamma_{\ell}(u,v)$ denotes a symmetric and nonnegative definite autocovariance for any $\ell$. In practice, the long-run covariance function can be estimated from a functional time series. Given its definition as a bi-infinite sum, a natural estimator of $C$ is
\begin{equation*}
\widehat{C}_{\varphi,\kappa}(u, v) = \sum^{\infty}_{\ell=-\infty}W_{\kappa}\left(\frac{\ell}{\varphi}\right)\widehat{\gamma}_{\ell}(u,v),
\end{equation*}
where $\varphi$ is known as the bandwidth parameter, and 
\begin{align*}
   \widehat{\gamma}_\ell(u, v)=\left\{
     \begin{array}{lr}
      \displaystyle \frac{1}{n}\sum_{t=1}^{n-\ell}\left[\X_t(u)-\overline{\X}(u)\right]\left[\X_{t+\ell}(v)-\overline{\X}(v)\right],\quad &\ell \ge 0
      \vspace{.3cm} \\
     \displaystyle \frac{1}{n}\sum_{t=1-\ell}^{n}\left[\X_t(u)-\overline{\X}(u)\right]\left[\X_{t+\ell}(v)-\overline{\X}(v)\right],\quad &\ell < 0,
     \end{array}
   \right.
\end{align*}
is an estimator of $\gamma_l(x,u)$, and $W_{\kappa}$ is a symmetric weight function with bounded support of order $\kappa$. The estimation accuracy of this estimator crucially depends on the bandwidth parameter, which can be selected by a plug-in algorithm of \cite{RS17}. With the estimated long-run covariance, a set of dynamic functional principal components and their associated scores may be obtained.

\subsection{Selection of the number of components}

Although it can be a research topic on its own, there are several approaches for selecting the number of retained terms $K$:
\begin{inparaenum}
\item[1)] scree plots or the fraction of variance explained by the first few functional principal components \citep{Chiou12};
\item[2)] pseudo-versions of Akaike information criterion and Bayesian information criterion \citep{FMW05};
\item[3)] predictive cross-validation leaving out one or more curves \citep{FMW05};
\item[4)] bootstrap methods \citep{HV06}; and 
\item[5)] eigenvalue ratio test \citep{AH13, LY12}.
\end{inparaenum}
In this article, we determine $K$ by a modified eigenvalue ratio criterion introduced in \cite{LRS19}. The estimated value of $K$ is determined as the integer minimizing ratios of two adjacent empirical eigenvalues given by
\begin{equation*}
\widehat{K} = \argmin_{1\leq k\leq k_{\max}}\left\{\frac{\widehat{\lambda}_{k+1}}{\widehat{\lambda}_k}\times \mathds{1}\Big(\frac{\widehat{\lambda}_k}{\widehat{\lambda}_1}\geq \theta\Big) + \mathds{1}\Big(\frac{\widehat{\lambda}_k}{\widehat{\lambda}_1}<\theta\Big)\right\},
\end{equation*}
where $k_{\max}$ is a pre-specified positive integer, $\theta$ is a pre-specified small positive number, and $\mathds{1}(\cdot)$ is the binary indicator function. When without prior information about a possible maximum of $K$, it is unproblematic to choose a relatively large $k_{\max}$, e.g., $k_{\max}=\#\left\{k\big|\widehat{\lambda}_{k}\geq \sum^{n}_{k=1}\widehat{\lambda}_{k}/n, k\geq 1\right\}$ \citep{AH13}. Given that the small empirical eigenvalues $\widehat{\lambda}_k$ for some $K<k<k_{\max}$ are close to zero, we adopt the threshold constant $\theta=1/\ln[\max(\widehat{\lambda}_1, n)]$ to ensure consistency of $\widehat{K}$.


\subsection{Univariate time series forecasting methods}

The forecasts of these principal component scores can be obtained via either a multivariate or univariate time series forecasting method. Among the multivariate time series forecasting methods, the vector autoregressive moving average model is commonly used \citep[see, e.g.,][]{ANH15}. However, these models often require stationarity and do not have an automatic order selection algorithm. In contrast, a univariate time series forecasting method, such as the autoregressive integrated moving average (ARIMA) model, can handle the non-stationarity of the principal component scores. Also, we use the automatic algorithm of \cite{HK08} to choose the optimal orders of autoregressive $p$, moving average $q$, and difference order $d$.

For each set of principal component scores, an ARIMA$(p, d, q)$ model is built, which has autoregressive components of order $p$, moving average components of order $q$, and degree of a difference $d$ needed to achieve stationarity. The model can be expressed as
\begin{equation*}
\Delta^d \beta_t = c + \sum^p_{l=1}\eta_{l}\Delta^d \beta_{t-l} + \pi_t + \sum^q_{\nu=1}\psi_{\nu}\pi_{t-\nu}, \qquad t=\max(p, q) + 1, \dots, n,
\end{equation*}
where $\Delta^d \beta_t$ represents the stationary time series after applying the difference operator of order $d$, $c$ is the drift term; $\{\eta_1,\dots,\eta_p\}$ represents the coefficients of the autoregressive components; $\{\psi_1,\dots,\psi_q\}$ represents the coefficients of the moving average components; and $\pi_t$ is a sequence of i.i.d. random variables with mean zero and a finite variance.

In the automatic ARIMA model, $d$ is selected based on successive Kwiatkowski-Phillips-Schmidt-Shin (KPSS) unit root tests \citep{KPSS92}. KPSS tests are used for testing the null hypothesis that an observable time series is stationary around a deterministic trend. We test the original data (i.e., the first set of principal component scores) for a unit root; if the test result is significant, we test the differenced data for a unit root. The procedure continues until we obtain our first insignificant result. After determining the value of $d$, the orders of $p$ and $q$ are selected based on the optimal Akaike information criterion with a correction for finite sample sizes \citep{Akaike74}. Having identified the optimal ARIMA$(p, d, q)$ model, the maximum likelihood method can estimate the parameters.

\subsection{Forecasting functional time series}

From the estimated long-run covariance function, we can extract estimated dynamic functional principal components $\bm{B} = \{\widehat{\phi}_1(u), \dots, \widehat{\phi}_K(u)\}$. Conditioning on the past functions $\bm{\X}(u)$ and the estimated functional principal components $\bm{B}$, the $h$-step-ahead point forecast of $\X_{n+h}(u)$ can be expressed as
\begin{equation*}
\widehat{\X}_{n+h|n}(u) = \text{E}\left[\X_{n+h}(u)|\bm{\X}(u), \bm{B}\right] =\overline{\X}(u) + \sum^K_{k=1}\widehat{\beta}_{n+h|n,k}\widehat{\phi}_k(u),
\end{equation*}
where $\widehat{\beta}_{n+h|n,k}$ denotes univariate time series forecasts of the $k$\textsuperscript{th} principal component scores and $h$ denotes a forecast horizon.

\begin{figure}[!ht]
\centering
\subfloat[Estimated mean function]
{\includegraphics[width=8cm]{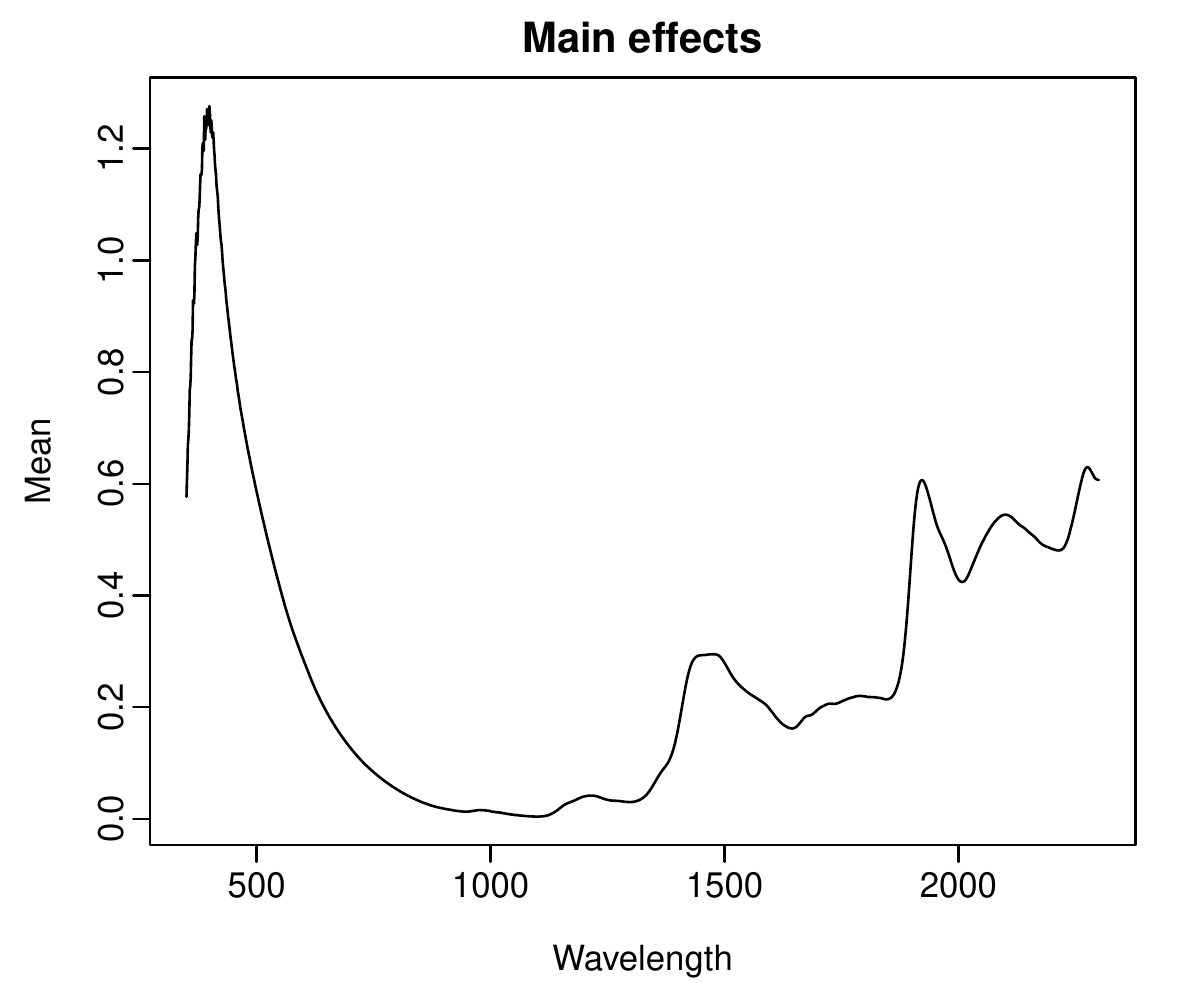}}
\quad
\subfloat[Estimated functional principal component $\widehat{\phi}_1(u)$]
{\includegraphics[width=8cm]{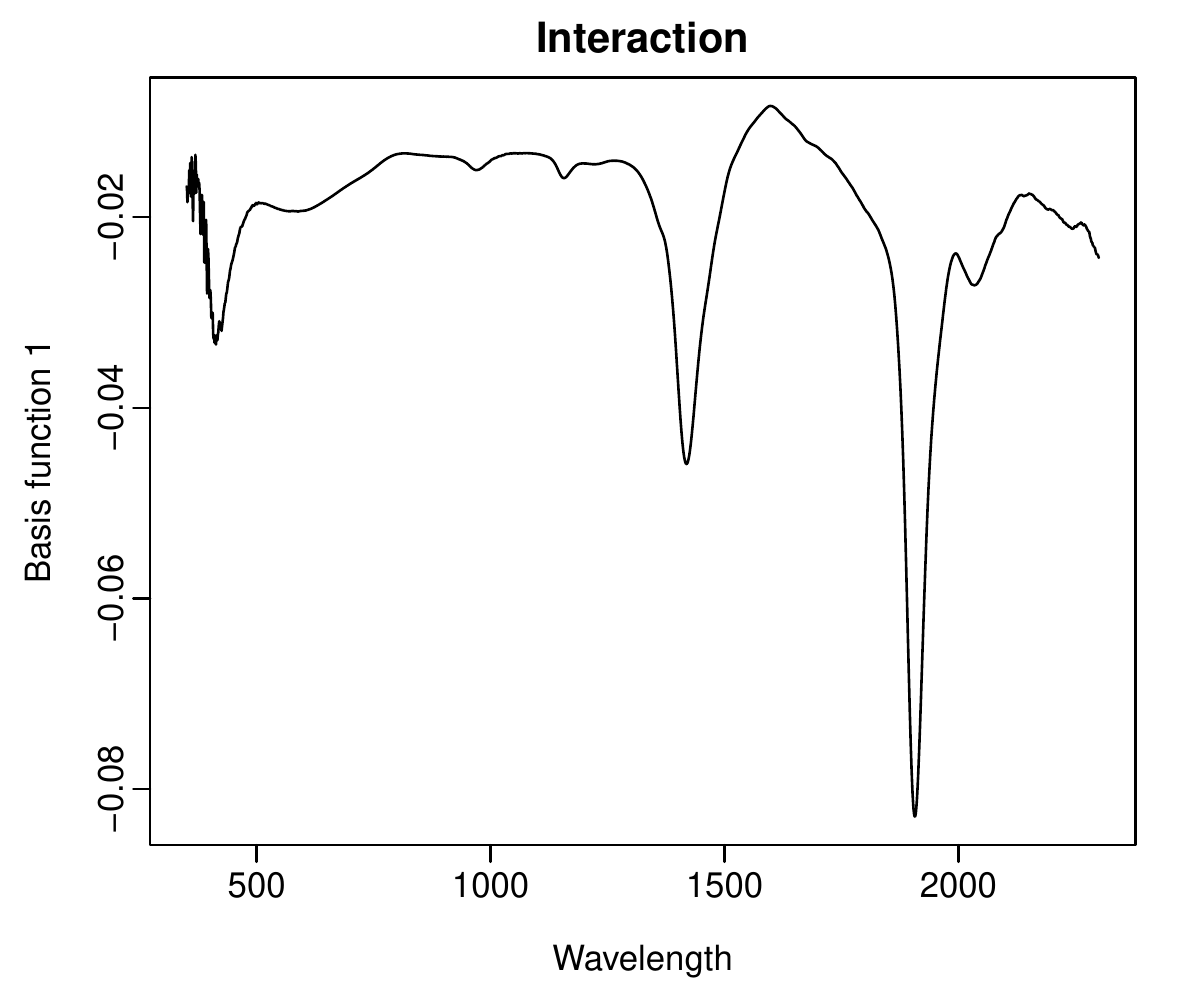}}
\\
\subfloat[First estimated principal component scores]
{\includegraphics[width=8cm]{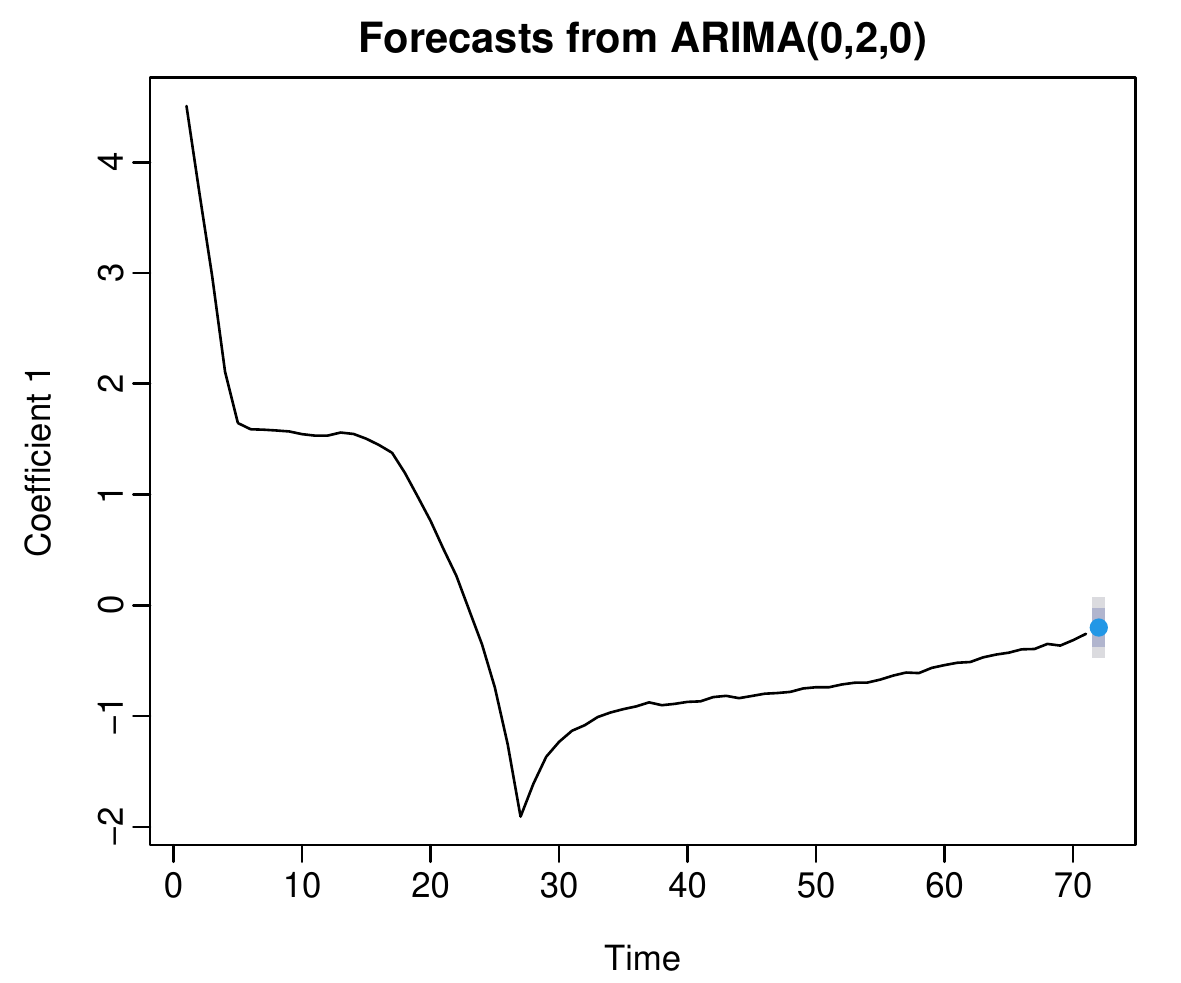}}
\quad
\subfloat[One-step-ahead forecast NIR curve]
{\includegraphics[width=8cm]{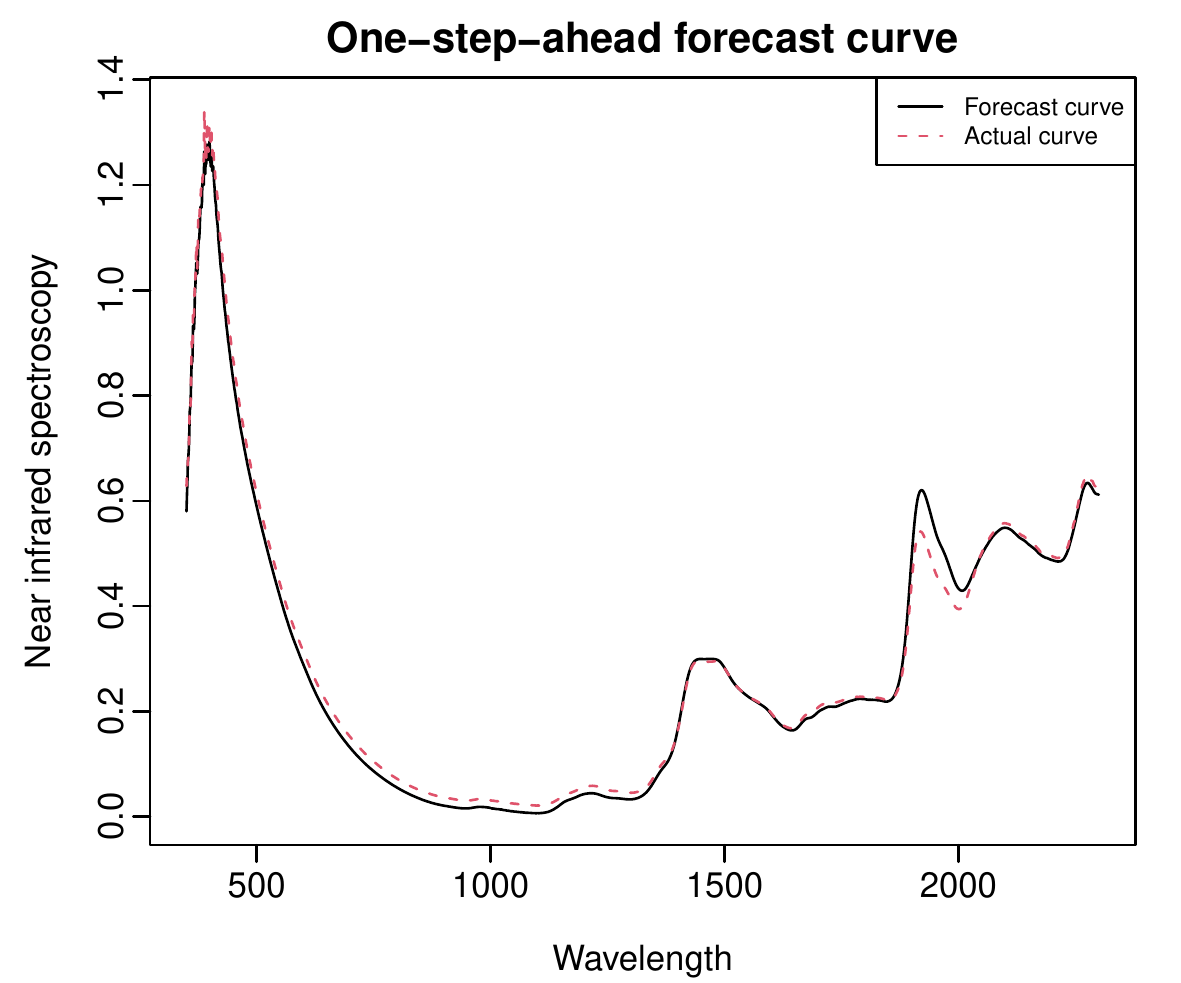}}
\caption{By applying a dynamic functional principal component analysis to the observed NIR spectra from time points 1 to 71, we obtain the estimated mean function, the first estimated functional principal component, and its associated scores. Using the eigenvalue ratio criterion, the number of retained components is one. By forecasting the scores with the ARIMA$(0,2,0)$ model, we obtain the one-step-ahead forecast curve conditional on the estimated functional principal component and mean function.}\label{fig:1_2}
\end{figure}

Let us consider our motivating example, consisting of 72 NIR spectra observed over time. Suppose we observe the first 71 NIR spectra, and we aim to produce a one-step-ahead curve forecast. Through a dynamic functional principal component analysis, we obtain estimated functional principal components and their associated scores in Figure~\ref{fig:1_2}.

\section{Stopping time detection}\label{sec:4}

\subsection{Rolling window scheme}\label{sec:4.1}

We consider a rolling window scheme of a functional time series forecasting method commonly used to assess model and parameter stabilities over time. It assesses the constancy of a model's parameter by computing parameter estimates and their corresponding forecasts over a rolling window of a fixed size through the entire sample. 

In the forecasting scheme, we produce one-step-ahead point forecasts using the first three functional observations initially. We re-estimate the parameters in a functional time-series model using the second to fourth functional observations. Forecasts from the estimated models are then produced for one-step-ahead. We iterate this process by increasing the sample size by one observation until the end of the sample period. This process produces 69 one-step-ahead forecasts. We compare these forecasts with the holdout samples (i.e., 4\textsuperscript{th} to 72\textsuperscript{th} curves) to determine the point forecast accuracy.


To evaluate the point forecast accuracy, we use the integrated squared forecast error (ISFE) \citep[see also][]{HU07}, which measures how close the forecasts are compared to the actual values of the variable being forecast. It can be written as
\begin{equation*}
\epsilon_{\gamma+1} = \int_{\mathcal{I}} \left[\X_{\gamma+1}(u) - \widehat{\X}_{\gamma+1|\gamma}(u)\right]^2du, \qquad \gamma = 3, \dots, (n-1),
\end{equation*}
where $\X_{\gamma+1}(u)$ represents the $(\gamma+1)$\textsuperscript{th} holdout sample in the forecasting scheme, while $\widehat{\X}_{\gamma+1|\gamma}(u)$ represents the one-step-ahead point forecasts for the holdout sample. 

\subsection{Regression-based approach}\label{sec:4.2}

Since our goal is to determine the optimal stopping time, the number of breakpoints is one. To estimate the breakpoint, we follow the methodology as set out in \cite{BP03} \citep[refer to][for a detailed description of an implementation]{ZKK+03, ZK05}. Suppose we observe a univariate time series of ISFEs, denoted by $\epsilon_{\gamma+1}$ for $\gamma = 3, \dots, (n-1)$. We estimate a random walk with a piecewise constant drift for the time-dependent variables:
\[ \Delta \epsilon_{\gamma+1} = \left\{ \begin{array}{ll}
         \varsigma_1 + \varepsilon_{\gamma+1} & \mbox{\qquad$\gamma+1 \leq \gamma^*$}\\
        \varsigma_{2}+\varepsilon_{\gamma+1} & \mbox{\qquad$\gamma+1 > \gamma^*$}\end{array} \right. \]
where $\varsigma_1$ and $\varsigma_2$ denote the respective mean terms before and after a breakpoint, and $\varepsilon_{\gamma+1}$ is the error term. We estimate this model using ordinary least squares by minimizing the sum of squared residuals (SSR):
\begin{equation*}
\text{SSR}\left(\gamma^*\right) = \sum_{\gamma=3}^{\gamma^*-1}\left(\Delta \epsilon_{\gamma+1} - \varsigma_{1}\right)^2 + \sum_{\gamma=\gamma^*}^{n-1}\left(\Delta \epsilon_{\gamma+1} - \varsigma_2\right)^2.
\end{equation*}

This model specification identifies one breakpoint, which divides the univariate time series of the forecast errors up into two regimes with different shifts. The optimal stopping time is determined as the one that minimizes the SSR. 

\section{Bootstrapping functional time series forecasts}\label{sec:5}

Bootstrapping has been receiving increasing attention in the functional time series literature as a way of quantifying uncertainty. \cite{PR94} obtained weak convergence results for approximate sums of weakly dependent, Hilbert space-valued random variables. \cite{DSW15} also obtained weak convergence results for Hilbert space-valued random variables, which are assumed to be weakly dependent in the sense of near-epoch dependence and showed consistency of a non-overlapping block bootstrap procedure. \cite{Nyarige16} and \cite{FN19} proposed a residual-based bootstrap for functional autoregressions. They showed that the empirical distribution of the centered sample innovations converges to the distribution of the innovations with respect to the Mallows metric. \cite{PPS19} established theoretical results for the moving block and the tapered block bootstrap. \cite{Shang18} applied a maximum entropy bootstrap procedure. \cite{Paparoditis18} considered the functional autoregressions and derived bootstrap consistency as the sample size and order of functional autoregression both tend to infinity. From a nonparametric viewpoint, \cite{FV11} applied a residual-based bootstrap procedure to construct confidence intervals for the regression function. \cite{ZP17} proposed a kernel estimation of the first-order nonparametric functional autoregression model and its bootstrap approximation.

Bootstrapping helps with statistical inference and forecast uncertainty quantification. \cite{PS21} proposed a sieve bootstrap for constructing prediction bands. Since the sieve bootstrap method can handle model misspecification, it improves the calibration of interval forecasts. However, the sieve bootstrap relies on stationarity in a functional time series, where our NIR data set does not hold. Instead, we consider a nonparametric bootstrap method of \cite{HS09}. In the principal component decomposition, at least three sources of uncertainty need to be considered. These are truncation errors in the principal component decomposition (concerning the estimated number of retained principal components), estimated mean function and functional principal components, and forecast errors in the forecast principal component scores. Since principal component scores are regarded as surrogates of the original functional time series, these principal component scores capture the temporal dependence structure inherited in the original functional time series \citep{Paparoditis18, Shang18b}. By adequately bootstrapping the forecast principal component scores, we can generate a set of bootstrapped curve forecasts conditional on the estimated mean function and estimated functional principal components from the observed functional time series.

Using a univariate time series forecasting method, we can obtain one-step-ahead forecasts for each set of the estimated principal component scores, $\{\widehat{\beta}_{1,k},\dots, \widehat{\beta}_{n,k}\}$ for $k=1,\dots,K$. Let the one-step-ahead forecast errors be given by $\widehat{\vartheta}_{t-1,k} = \widehat{\beta}_{t,k}-\widehat{\beta}_{t|t-1,k}$ for $t=2,\dots,\gamma$ and $\gamma = 3,\dots,(n-1)$. These can then be sampled with replacement to give a bootstrap sample of $\beta_{\gamma+1,k}$:
\begin{equation*}
\widehat{\beta}^{(b)}_{\gamma+1|\gamma,k}=\widehat{\beta}_{\gamma+1|\gamma,k}+\widehat{\vartheta}_{*,k}^{(b)},\qquad b=1,\dots,B,
\end{equation*}
where $B=1,000$ symbolizes the number of bootstrap replications and $\widehat{\vartheta}_{*,k}^{(b)}$ are sampled with replacement from
$\{\widehat{\vartheta}_{1,k},\dots,\widehat{\vartheta}_{\gamma-1,k}\}$.

Assuming the first $K$ principal components approximate the original functional time series relatively well, the model residual should contribute nothing but random noise. Consequently, we can bootstrap the model residuals in~\eqref{eq:FPCA_finite} by sampling with replacement from the model residual term $\{\widehat{\zeta}_1(u),\dots, \widehat{\zeta}_{\gamma-1}(u)\}$.

Adding two components of variability, we obtain $B$ variants for $\X_{\gamma+1}(u)$,
\begin{equation*}
\widehat{\X}_{\gamma+1|\gamma}^{(b)}(u) = \sum^K_{k=1}\widehat{\beta}_{\gamma+1|\gamma,k}^{(b)}\widehat{\phi}_{k}(u)+\widehat{\zeta}^{(b)}_{\gamma+1}(u).
\end{equation*}
With the bootstrapped $\{\widehat{\X}^{(1)}_{\gamma+1|\gamma}(u),\dots, \widehat{\X}^{(B)}_{\gamma+1|\gamma}(u)\}$, we fit a functional time series model in Section~\ref{sec:3}, where the retained number of principal components is estimated from eigenvalue ratio criterion and is allowed to be different in bootstrap samples. We obtain a one-step-ahead forecast by conditioning the estimated mean function and functional principal components in each bootstrap sample. Then, we compute the one-step-ahead bootstrap forecast errors between the bootstrapped forecasts and actual holdout samples,
\begin{equation*}
\epsilon_{\gamma+1}^{(b)} = \int_{\mathcal{I}}\left[\X_{\gamma+1}(u) -\widehat{\X}^{(b)}_{\gamma+1|\gamma}(u)\right]^2du.
\end{equation*}
With each bootstrapped errors $\{\epsilon_4^{(b)},\dots,\epsilon_n^{(b)}\}$, we apply the structure break detection algorithm in Section~\ref{sec:4.2} to find the estimates of the optimal stopping time.

\section{Simulation studies}\label{sec:6}

We consider two data generating processes. In Section~\ref{sec:6.1}, we consider a stationary functional time series process, where there is a presence of abrupt change at a pre-fixed location. In Section~\ref{sec:6.2} and~\ref{sec:6.3}, we consider a non-stationary functional time series process, where there is a presence of abrupt and gradual change at a randomly assigned location, respectively.

\subsection{An abrupt change in the mean of a stationary functional time series}\label{sec:6.1}

\begin{figure}[!htbp]
\centering
\includegraphics[width=11cm]{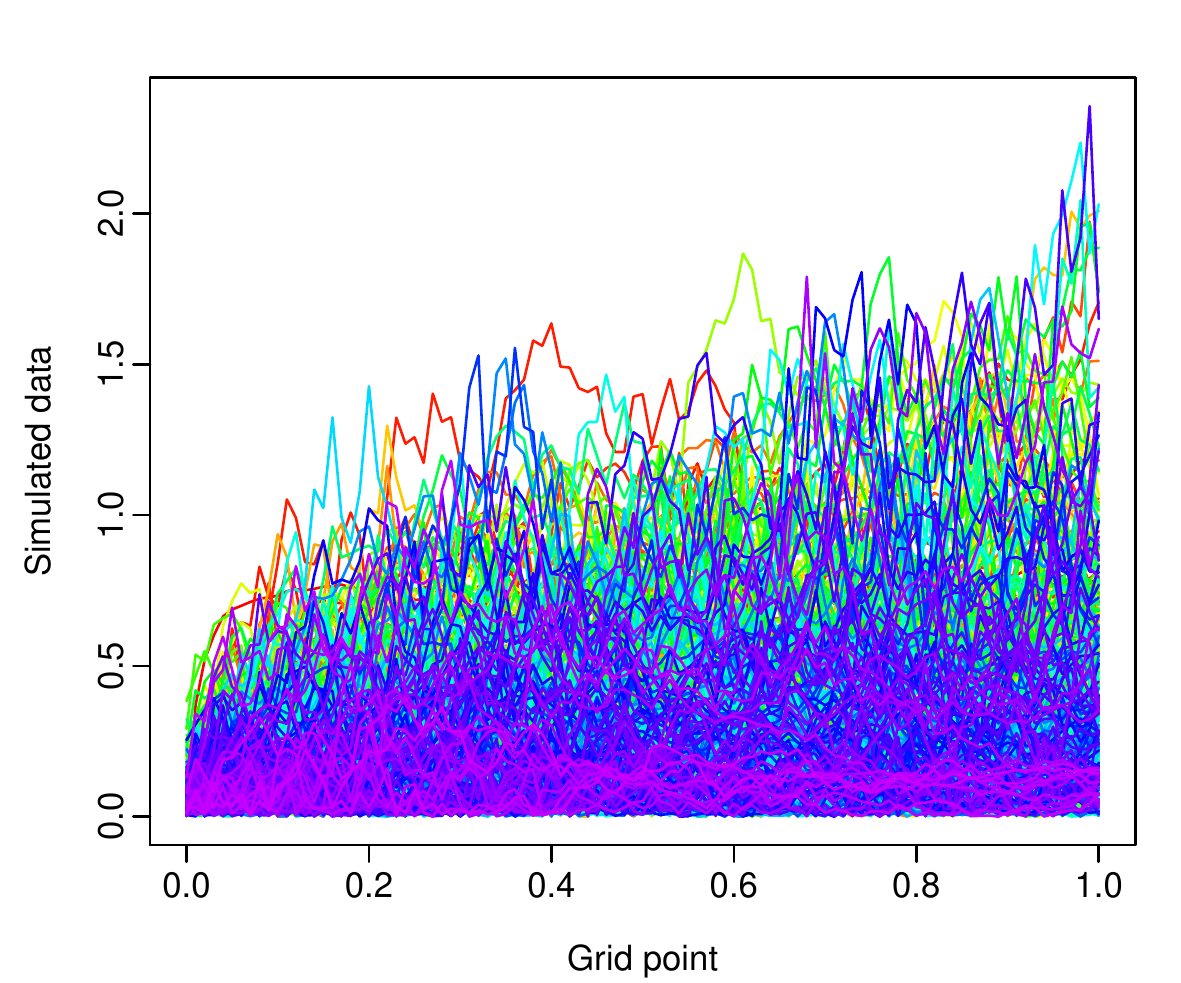}
\caption{One replication of $n=400$ simulated curves $\{\mathcal{Y}_1(u),\dots,\mathcal{Y}_{400}(u)\}$.}\label{fig:2}
\end{figure}

We utilize Monte Carlo simulation to evaluate the performance of our method. The data generating process is a pointwise FAR$(1)$, given by
\begin{align*}
\X_1(u) &= 10\times u \times (1-u) + \omega \times B_1(u) \\
\X_t(u) &= (\rho + c) \X_{t-1}(u) + \omega \times B_t(u), \qquad t = 2,\dots,n \\
\mathcal{Y}_t(u) &= \frac{|\X_{t-1}(u) - \X_{t}(u)|}{|\X_{t-1}(u) + 0.1|},
\end{align*}
where $\{B_t(u), t=1,\dots,n, u\in [0,1]\}$ denote i.i.d. standard Brownian motions. In practice, we discretize continuum $u$ on 101 equally spaced grid points. We consider three values of $\omega = 0.1, 0.5, 0.9$ to reflect three levels of noise to signal. The coefficients satisfy $|\rho|<1$ and $|\rho+c|<1$ in order to ensure the stationarity. Let $\rho=0.2$ and $c = 0$ for curves from two to $\tau = \lceil n/2\rceil$; while $c=0.7$ for curves from $\lceil n/2\rceil+1$ to $n$. An example of simulated curves with a sample size $n=400$ is presented in Figure~\ref{fig:2}, where the actual stopping time is 200.

\begin{figure}[!htbp]
\centering
\subfloat[$\omega = 0.1$]
{\includegraphics[width=5.5cm]{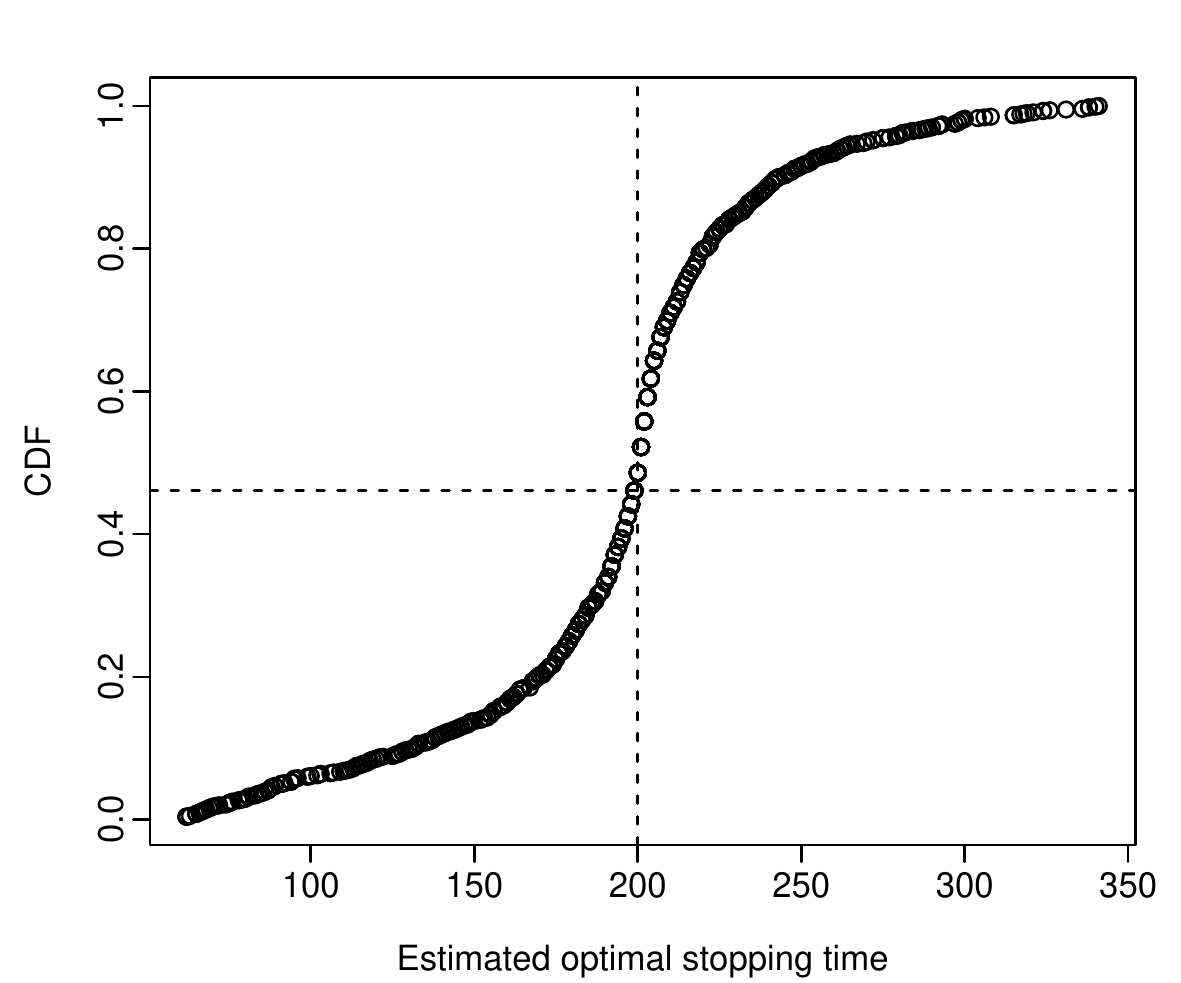}}
\subfloat[$\omega = 0.5$]
{\includegraphics[width=5.5cm]{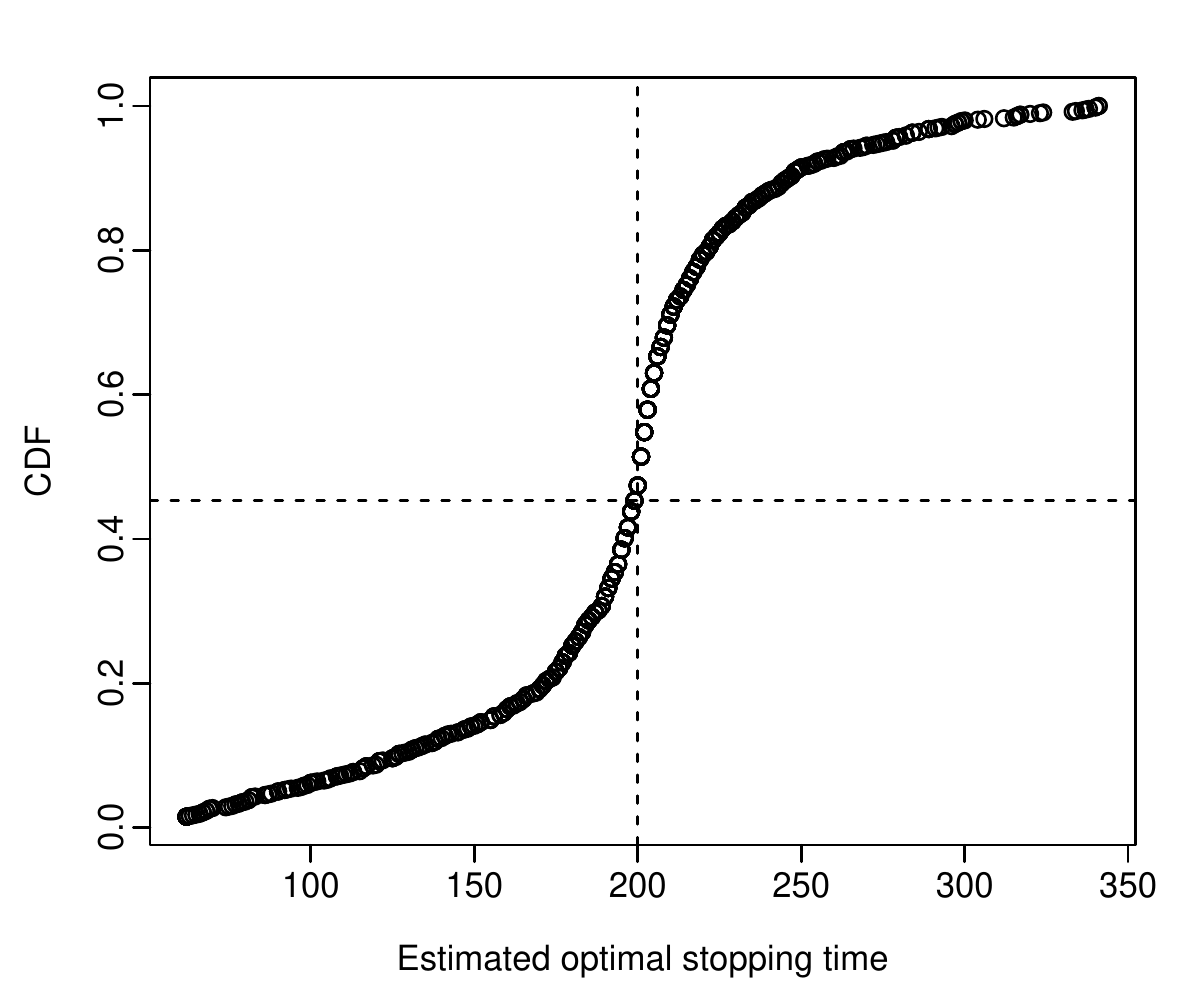}}
\subfloat[$\omega = 0.9$]
{\includegraphics[width=5.5cm]{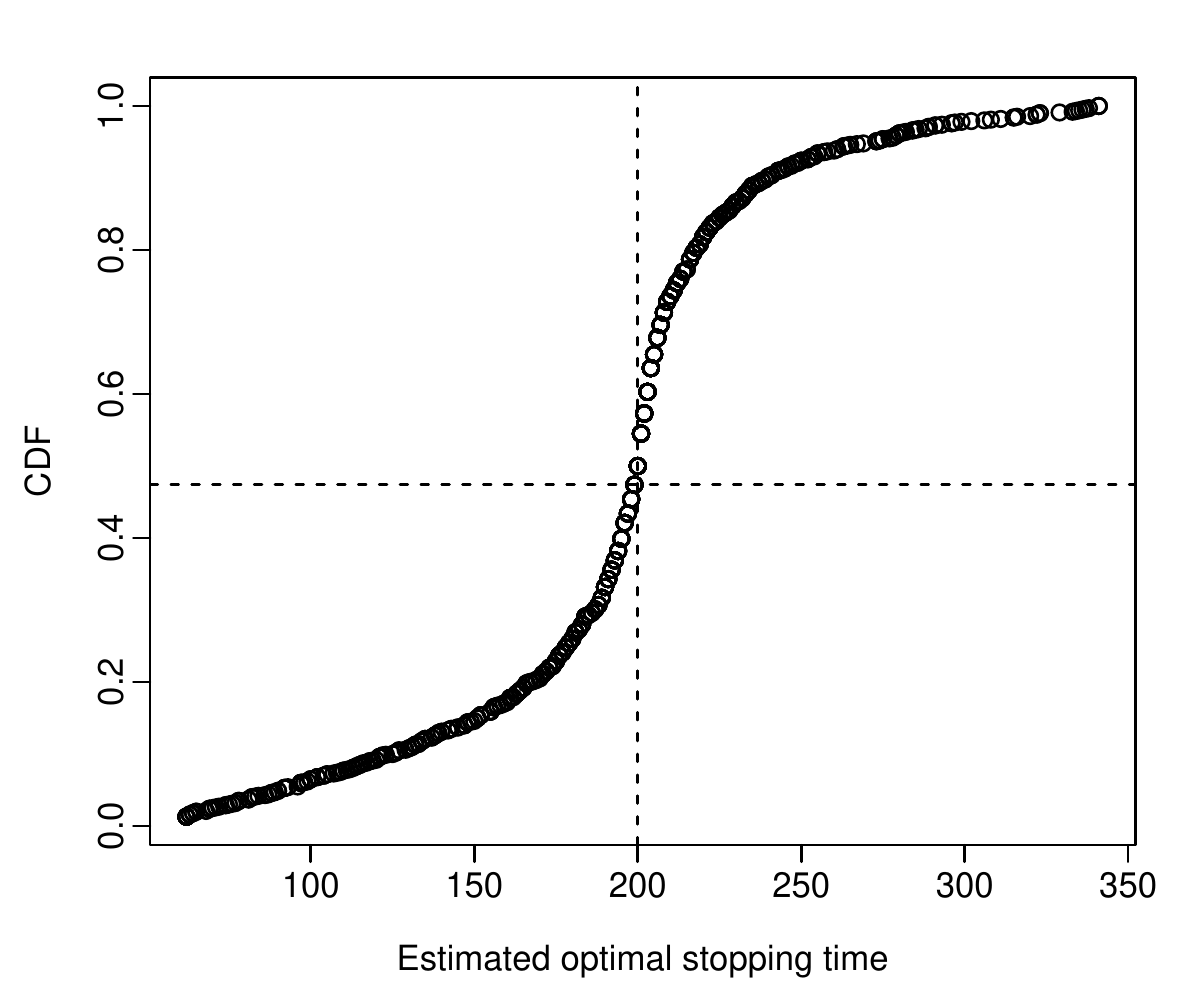}}
\\
\caption{With 1,000 replications, we compute the empirical CDFs of the stopping time using the regression-based approach with a sample size of $n=400$ for a noise-to-signal ratio $\omega=0.1, 0.5, 0.9$.}\label{fig:3}
\end{figure}

In each replication, we apply our regression-based approach to estimating the optimal stopping time. With the estimated stopping time points from 1,000 replications, we present cumulative distribution functions (CDFs) for the regression-based approach with three different noise to signal levels in Figure~\ref{fig:3}.

In Table~\ref{tab:1}, we tabulate the number of times among 1,000 replications when the estimated stopping time is greater than or equal to the actual stopping time. As sample size $n$ increases, the chance of covering the holdout stopping time generally decreases, but the median of the estimated stopping time becomes more accurate. For the three $\omega$ values, the possibility of covering the holdout stopping time generally decreases from $\omega=0.1$ to $\omega=0.9$. The smaller the $\omega$ value is, the larger the signal-to-noise ratio is. From Table~\ref{tab:1}, the mean and median of the estimated stopping time are not distant from the actual stopping time. 

\begin{table}[!htbp]
\tabcolsep 0.36in
\centering
\caption{For three sample sizes, we determine the mean and median of the estimated change points and the number of times out of 1,000 replications when the estimated stopping time is greater than or equal to the actual stopping time. From a conservative aspect, it is essential not to miss the actual stopping time. Out of 1,000 replications, each number in the table reflects a probability where the actual stopping time is within the estimated stopping time.}\label{tab:1}
\begin{tabular}{@{}lllrrr@{}}
\toprule
$n$ & $\tau$ & $\omega$ & Mean of $\widehat{\tau}$  & Median of $\widehat{\tau}$ & $\#$($\widehat{\tau}$ $>=$ $\tau$)\\
\midrule
101 & 51 & 0.1 & 51.965 & 52& 572 \\
&&  0.5 & 51.497 & 53 & 567 \\
&&  0.9 & 50.822 & 52 & 546 \\
\\
201 & 101 & 0.1 & 100.079 & 102 & 528 \\
&&  0.5 & 98.009 & 101 & 523 \\
&& 0.9 & 97.742 & 101 & 515 \\
\\
401 & 201 & 0.1 & 195.091 & 201 & 514 \\
&&  0.5 & 195.464 & 201 & 526 \\
&& 0.9 & 193.313 & 200.5 & 500 \\
\midrule
\end{tabular}
\end{table}

\subsection{An abrupt change in the mean of a non-stationary functional time series}\label{sec:6.2}

We consider another data generating process for simulating functional time series. We begin with simulating a time series of error functions $[\epsilon_1(u), \epsilon_2(u), \dots, \epsilon_n(u)]$ given below:
\begin{equation}
\epsilon_t(u) = \sum^K_{k=1}\beta_{t,k}\phi_k(u) + \zeta_t(u),\label{eq:Aue_1}
\end{equation}
where $\left[\phi_1(u), \phi_2(u), \dots, \phi_K(u)\right]$ are randomly sampled with replacement from $K=21$ Fourier basis functions, and $\zeta_t(u)$ denotes innovation term that can be independent over $t$. We consider 101 equally-spaced grid between 0 and 1.

Let $\bm{\beta}_t = (\beta_{t,1}, \beta_{t,2},\dots,\beta_{t,K})^{\top}$ be a $K$-dimensional vector. We generate $\bm{\beta}_t$ from a vector autoregression of order 1 (VAR(1)) model, 
\begin{equation*}
\bm{\beta}_{t} = \bm{A}\bm{\beta}_{t-1} + \bm{\xi}_{t},\qquad t=2,\dots,n,
\end{equation*}
where $\bm{A} = (a_{ij})_{K\times K}$ is the VAR(1) coefficient matrix, and $\bm{\xi}_t$ denotes the error term of the VAR(1) model at time $t$. Following \cite{LRS19}, we consider two possible structures for $\bm{A}$:
\begin{inparaenum}
\item[1)] a diagonal matrix with diagonal elements drawn from a $U(-0.5, 0.5)$ and $\bm{\xi}_{t}$ is  generated by a $K$-dimensional normal distribution with mean zero and power-decay covariance structure
\begin{equation*}
\text{cor}(\xi_t^i, \xi_t^j) = \rho^{|i-j|},
\end{equation*}
where $\rho$ denotes a correlation parameter, such as $\rho=0.5$.
\item[2)] Alternatively, $\bm{A}$ is a banded autoregressive matrix with $a_{i,j}$ indpendently drawn from a $U(-0.3, 0.3)$ when $|i-j|\leq 3$ and $a_{ij}=0$ when $|i-j|>3$, and $\bm{\xi}_{t}$ is independently generated by a $K$-dimensional normal distribution with mean zero and identity covariance matrix.
\end{inparaenum}

To specify a change-point location for the population, we draw a value from a $U(0.25\times n, 0.75 \times n)$. The lower and upper bounds of the uniform distribution are purposely chosen so that the location of a change point does not lie on the boundary of a sample. We could divide the change-point location $\tau$ by sample size $n$ to compute a probability $\mathfrak{p}\in (0.25, 0.75)$.

Following \cite{ARS18}, a class of break functions was given by
\begin{align*}
\delta_k^*(u) &= \frac{1}{\sqrt{k}}\sum^k_{w=1}\phi_{w}(u),\qquad k=1,2,\dots,K, \\
\delta_k(u) &= \delta_k^*(u) \times \sqrt{c},
\end{align*}
where the normalization is required to ensure $\delta_k^*(u)$ has unit norm. For a population, $\delta_1(u)$ is the case of a break only in the leading eigendirection, while $\delta_K(u)$ is the case of a break that affects all eigendirections \citep[see, e.g.,][]{ARS18}. The value of $c$ controls the magnitude of the break, and it links to the signal-to-noise ratio
\begin{equation}
\text{SNR} = c\times \frac{\mathfrak{p}(1-\mathfrak{p})}{\text{tr}(\widehat{C}_{\epsilon})},
\end{equation}
where $\text{tr}(\widehat{C}_{\epsilon})$ denote the trace of the estimated long-run covariance of the error term in~\eqref{eq:Aue_1}. We can compute the value of $c$ for a given SNR value.

With a chosen eigendirection $k$, such as $k=1$, we simulate $n$ samples of a non-stationary functional time series as follows:
\begin{align*}
\X_t(u) &= \delta_k(u)\times \mathds{1}\{t>\tau\} + \epsilon_t(u), \\
\Y_t(u) &= \overline{\X}(u) + \X_t(u),
\end{align*}
where $\overline{\X}(u) = \frac{1}{n}\sum^n_{t=1}\X_t(u)$.

In Table~\ref{tab:DGP_2}, we present some summary statistics of our estimated change points for three sample sizes and two SNRs. As SNR increases from 0.01 to 0.1, the mean and median of the estimated change points are generally closer to those of the actual change points for the same sample size. The probability that the actual stopping time is within the estimated stopping time also increases.

\begin{table}[!htbp]
\centering
\tabcolsep 0.07in
\caption{For three sample sizes, we determine the mean and median of the actual and estimated change points and the number of times out of 1,000 replications when the estimated stopping time is greater than or equal to the actual stopping time.}\label{tab:DGP_2}
\begin{tabular}{@{}llrrrrrrrr@{}}
\toprule
& & \multicolumn{2}{c}{True change point $\tau$} & \multicolumn{6}{c}{Estimated change point $\widehat{\tau}$} \\
\cmidrule{3-10}
& & 	& & \multicolumn{3}{c}{$\text{SNR}=0.1$} & \multicolumn{3}{c}{$\text{SNR}=0.01$} \\
\cmidrule{5-10}
$\bm{A}$  & $n$ & Mean & Median & Mean & Median & $\#$($\widehat{\tau}$ $>=$ $\tau$) & Mean & Median & $\#$($\widehat{\tau}$ $>=$ $\tau$) \\
\midrule
band  & 100 &  50.07 & 50 & 50.23 & 50 & 29 & 51.60 & 52  & 393 \\
		& 200 & 99.99 & 99 & 102.08 & 102.5 & 67 & 107.72 & 108 & 218	\\
		& 400 & 200.83 & 203.5 & 203.67 & 210.5 & 150 & 186.79 & 177 & 407	\\
\\
diag  & 100 & 49.46 & 49 & 50.99 & 50.5 & 47 & 51.37 & 51 & 356 \\	
	 	& 200 & 99.60 & 100 & 101.64 & 102.5 & 84 & 102.94 & 106.5 & 313	\\
		& 400 & 200.39 & 199.5 & 202.72 & 205 & 168 & 198.45 & 200 & 395	\\
\bottomrule
\end{tabular}
\end{table}

\subsection{A gradual change in the mean of a non-stationary functional time series}\label{sec:6.3}

While Sections~\ref{sec:6.1} and~\ref{sec:6.2} present stationary and non-stationary functional time series processes with an abrupt change in mean, in this section, we modify the data generating process of Section~\ref{sec:6.2} from an abrupt change to a gradual change in mean. With a chosen eigendirection $k$, such as $k=1$, we simulate $n$ samples of a non-stationary functional time series as follows:
\begin{align*}
\X_t(u) &= \sqrt{t} \times \frac{n^{\alpha}}{\sqrt{n}} \times \delta_k(u) \times \mathds{1}\{t>\tau\}+\epsilon_t(u), \\
\Y_t(u) &= \overline{\X}(u) + \X_t(u),
\end{align*}
where $t$ is a time index representing a gradual change. Also, $\alpha\in (0, \frac{1}{2})$ is a constant, and together with $c$ control the magnitude of the change point. As $\alpha$ or $c$ increases, the magnitude of the change point increases; thus, it is easier to be detected.
 
In Table~\ref{tab:DGP_3}, we present some summary statistics of our estimated change points for three sample sizes and two SNRs. As $\alpha$ increases for the same SNR and sample size, the mean and median of the estimated change points are closer to those of the actual change points. As SNR increases from 0.01 to 0.1, the mean and median of the estimated change points are closer to those of the actual change points for the same sample size and same $\alpha$ value. 
 
\begin{table}[!htbp]
\centering
\tabcolsep 0.075in
\caption{We determine the mean and median of the estimated and actual change points from 1,000 replications for three sample sizes and two SNRs.}\label{tab:DGP_3}
\begin{tabular}{@{}lllrrrrrrrr@{}}
\toprule
& & & \multicolumn{2}{c}{True change point $\tau$}   & \multicolumn{6}{c}{Estimated change point $\widehat{\tau}$} \\
\cmidrule{4-11}
& & & & & \multicolumn{2}{c}{$\alpha=0.05$} & \multicolumn{2}{c}{$\alpha=0.25$} & \multicolumn{2}{c}{$\alpha=0.45$} \\
\cmidrule{6-11}
$\bm{A}$ & SNR & $n$ & Mean & Median & Mean & Median & Mean & Median & Mean & Median  \\
\midrule
band & 0.1 & 100 &  50.07 & 50 & 51.62 & 52 & 51.55 & 51 & 51.54 & 51 \\
& & 200  & 99.99 & 99 & 102.87 & 103 & 101.56 & 102 & 101.51 & 101 \\
& & 400  & 200.83 & 203.5 & 205.07 & 211 & 202.31 & 205 & 202.28 & 203.5\\
\\
&  0.01 & 100 &&  & 52.20 & 54 & 51.72 & 52 & 51.55 & 52 \\
& & 200			& & & 110.14 & 113 & 103.55 & 104 & 101.58 & 102 \\
& & 400			& & & 191.58 & 187.5 & 204.61 & 211 & 202.30 & 205 \\
\\
diag & 0.1 & 100 & 49.46 & 49 & 51.12 & 51 & 51.00 & 51 & 50.99 & 51 \\
	& & 200  & 99.6 & 100 & 102.28 & 103 & 101.13 & 101 & 101.09 & 101 \\
	& & 400  & 200.39 & 199.5 & 203.82 & 206 & 201.91 & 202 & 201.89 & 201 \\
\\
& 0.01 & 100 & & & 52.61 & 54 & 51.33 & 52 &	51.00 & 51 \\
& & 200  & & & 105.33 & 111 & 102.60 & 103 & 101.13 & 101.5 \\
& & 400  & & & 204.27 & 215.5 & 203.80 & 206 & 201.90 & 201.5 \\ 
\bottomrule
\end{tabular}
\end{table}

\section{Application to the wood panel data set}\label{sec:7}

According to \cite{PGJ14}, the glue curing process is affected by the moisture content in a wood panel and process factors such as applied pressure and temperature. For better determination of phenolic glue curing of wood panels, FPInnovations exploited the near-infrared (NIR) spectroscopy to extract spectral signals that can reflect chemical changes due to the curing of the resin. In this study, these 72 spectra were obtained from a piece of Spectralon reference materials coated with a layer of urea-formaldehyde resin. This sample was placed in an oven to simulate the online process curing temperature and withdrawn at a periodic time for spectral measurements. Since the traditional glue manufacturing process can hardly be monitored, considerable energies must be consumed to guarantee the resin's curing completion. As a result, this new process monitoring tool ensures substantial savings in energy and time.

\begin{figure}[!htbp]
\centering
\includegraphics[width=10cm]{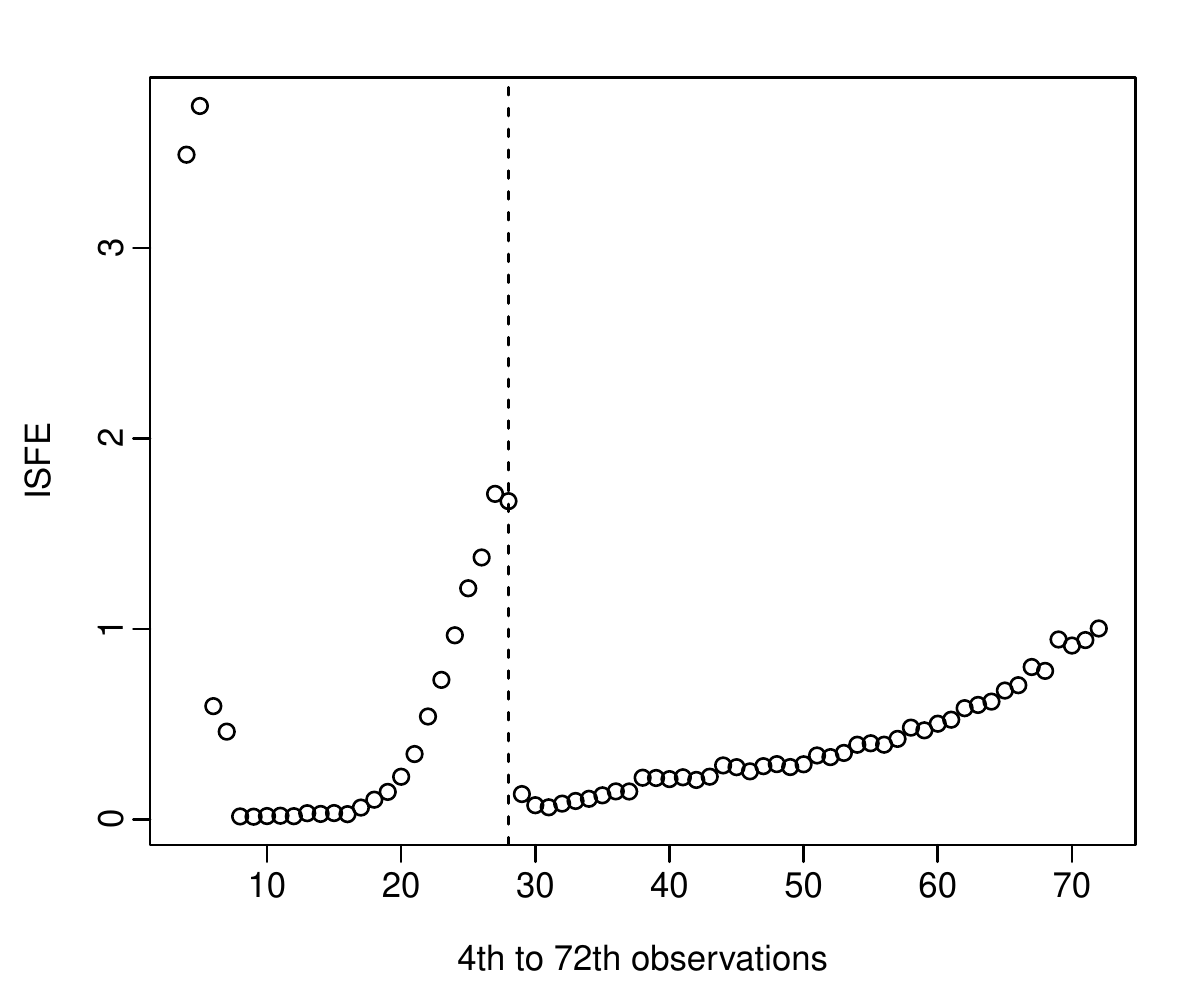}
\caption{Based on the ISFEs, we determine the estimated stopping time using the regression-based approach.}\label{fig:5}
\end{figure}

Using the eigenvalue ratio criterion, the number of retained components is determined to be one. With the functional time series forecasting method, we compute the ISFEs in Figure~\ref{fig:5}. From a univariate time series of these ISFEs, we then implement a structural change detection method described in Section~\ref{sec:4.2} to identify one breakpoint. This breakpoint is also our estimated optimal stopping time. The breakpoint occurs at time point 27 using the rolling window scheme. As a result, time point 27 may be the optimal stopping time using the regression-based approach.

The point estimate of the optimal stopping time is limited without an adequate assessment of probabilistic uncertainty associated with the point estimate. In computing the forecast errors, forecast uncertainty stems from systematic deviations (e.g., due to parameter and model uncertainties) and random fluctuations (e.g., due to the model error term). Thus, it is essential to provide an interval forecast as well as a point forecast of the optimal stopping time to
\begin{inparaenum}[(1)]
\item assess the future uncertainty level;
\item enable different strategies to be planned for a range of possible outcomes indicated by the interval forecast;
\item compare forecasts from different methods more thoroughly; and
\item explore different scenarios based on various assumptions.
\end{inparaenum}

The uncertainty associated with our estimated stopping time may vary largely due to the forecast errors in the forecasting schemes. We apply a functional time series forecasting method that utilizes the principal components and their associated scores to compute these forecast errors. By bootstrapping the scores and model residuals, we obtain a set of bootstrap functional time series conditional on the estimated mean function, estimated functional principal components, and observed functional time series. We compute the ISFEs and determine the optimal stopping time for each bootstrap replication using the bootstrap series. In turn, bootstrapping provides us with a distribution of the estimated optimal stopping time, from which we can determine the mode (frequently occurring value). The mode is time point 29 using the regression-based approach. Out of 1,000 bootstrap replications, there are only seven times where the stopping time is time point 30.

The detected time points 29 and 30 obtained from the 1,000 bootstrap samples differ from the time point 27 identified by the same approach applied to the original functional time series. An explanation is that the original functional time series are some realizations of an unknown stochastic process. The bootstrap samples generally mimic the temporal dependence exhibited in the original functional time series, but the bootstrap samples contain a random noise with a practically unknown signal-to-noise ratio. In turn, the detected time points may differ from the one from the original data.


The economic benefit of this empirical application is substantial. Among 72 time series of NIR spectra, our estimated optimal time point indicates that we do not need to observe the entire sample to uncover when the glue curing process may be completed. In turn, this finding could save energy, time, and labor costs.

For comparison, we also implemented a structural break method of \cite{ARS18} and detected the change point at time point 22 using both fully-function and functional principal component analysis-based approaches. The optimal stopping time is recommended at time point 30 for this data set. It is better to overestimate than underestimate the stopping time from an applied perspective since the process may not be completed if stopped too early.

\section{Conclusion}\label{sec:8}

We provide a solution for determining the optimal stopping time for phenolic glue curing of wood panels in an automated process environment from a time series of NIR spectra. This solution, based on the NIR spectroscopy technology, provides a novel alternative to monitoring the process. With a more accurate evaluation of the curing process's stopping time, the proposed solution leads to substantial savings in energy, time, and labor costs, thus enhancing the overall competitiveness of the Canadian wood industry. Furthermore, the proposed methodology can be applied to other applications such as qualitative analysis and quality control, where spectral signals can be employed to analyze a dynamical process of interest.

The essence of our methodology is to identify breakpoints by applying a structural change method to iterative one-step-ahead forecast errors. These forecast errors are obtained by computing the ISFEs between the holdout functional time series and their one-step-ahead forecasts. Our approach is general enough to apply other structural break methods and functional time series forecasting methods.

There are at least three ways in which the present paper can be extended: 
\begin{inparaenum}
\item[1)] We implement a regression-based approach to detect a single change point due to the nature of our problem. However, one could apply other change-point methods, including those that can detect multiple change points \citep[see, e.g.,][]{QWX19, WZ20, XC21}. Then, the stopping time is estimated by the point that has the longest homogenous segment to the end of the time points.
\item[2)] A challenging and important research direction is the change-point detection for high-dimensional functional time series. In this study, we observe only one time series of near-infrared spectroscopy curves. As the experimental conditions can change from one to another, the stopping time can vary among different wood products. If we observe multiple (high-dimensional) time series of such functional time series for various wood products, it would be helpful to propose a hypothesis test to check if there is a change point in any of the functional time series. If so, we may develop a change-point detection method to identify those functional time series that have a change point.
\item[3)] By bridging a gap between academia and industry partners, our change-point detection method can be applied by employees in the wood panel industry to verify the effectiveness of the stopping-time estimation for a range of wood products. We aim to enrich the collaboration with a developed computer code in \Rlogo.
\end{inparaenum}



\newpage
\bibliographystyle{agsm}
\bibliography{lumber}

\end{document}